\documentclass[preprint,12pt]{elsart}
\usepackage{graphicx,epsfig,amssymb,subfigure}

\newcommand{\sect}[1]{\setcounter{equation}{0}\section{#1}}
\newcommand{\subsect}[1]{\subsection{#1}}
\renewcommand{\theequation}{\thesection.\arabic{equation}}

\newtheorem{defi}{Definition}
\newtheorem{theo}{Theorem}
\newtheorem{exam}{{\large \bf Example}}[section]
\newcommand{\beq}{\begin{equation}}
\newcommand{\eeq}{\end{equation}}
\newcommand{\benum}{\begin{enumerate}}
\newcommand{\eenum}{\end{enumerate}}
\newcommand{\beqn}{\begin{eqnarray}}
\newcommand{\eeqn}{\end{eqnarray}}
\newcommand{\beqnn}{\begin{eqnarray*}}
\newcommand{\eeqnn}{\end{eqnarray*}}

\newcommand{\pa}{\partial}

\newcommand{\calA}{{\mathcal{A}}}

\newcommand{\calF}{{\mathcal{F}}}

\newcommand{\calH}{{\mathcal{H}}}

\newcommand{\calL}{{\mathcal{L}}}

\newcommand{\calV}{{\mathcal{V}}}

\begin{document}   

\begin{frontmatter}

\title{Recursive properties of Dirac and Metriplectic Dirac brackets with Applications}
\author[hn]{Sonnet Hung Q. Nguyen}, 
\ead{ hungnq\_kvl@vnu.edu.vn,  sonnet@impan.gov.pl }
\author[wa]{ {\L}ukasz A. Turski}
\ead{ laturski@cft.edu.pl}

\address[hn]{ 
         Hanoi University of Science, Nguyen Trai 334, Hanoi, Vietnam.}

\address[wa]{ 
          Center for Theoretical Physics, Polish Academy of Sciences,
          and Department of Mathematics and Natural Sciences,
          Cardinal Wyszynski University, 
          Al. Lotnik{\' o}w 32/46, 02-668 Warsaw, Poland }

\begin{abstract}

  In this article, we prove that Dirac brackets for Hamiltonian and
non-Hamiltonian constrained systems can be derived recursively.
  We then study the applicability of that formulation in analysis of some
interesting physical models. Particular attention is paid to feasibility of
implementation code for Dirac brackets in Computer Algebra System and
analytical techniques for inversion of triangular matrices.

\end{abstract}


\end{frontmatter}

\sect{Introduction}

The fundamental notion in the Hamiltonian formulation of classical dynamics
of particles and fields is the canonical Poisson bracket defined over the
space of all differentiable functions of the phase space (of even dimension),
such that: for each two phase space functions $f(q,p)$ and $g(q,p)$ where
$(q,p)=(q_1,\ldots,q_n,p_1,\ldots,p_n)$ denote generalized positions and
momenta respectively,
\beqn
    \{f,g\}=\frac{\partial f}{\partial q} \frac{\partial g}{\partial p} -
            \frac{\partial f}{\partial p} \frac{\partial g}{\partial q} =
    \sum_{k=1}^{n} \frac{\partial f}{\partial q_k} \frac{\partial g}{\partial p_k} -
    \frac{\partial f}{\partial p_k} \frac{\partial g}{\partial q_k} \,.
\eeqn
This bracket is linear in each argument, skew-symmetric: $\{f,g\}=-\{g,f\}$,
satisfies {\it Leibniz} identity:
$\{f, g \cdot h\} = \{f, g\} \cdot h + g \cdot \{f, h\}$,
{\it Jacobi} identity: $\{f,\{g,h\}\}+\{g,\{h,f\}\}+\{h,\{f,g\}\}=0$ and
is non-degenerate, i.e.
$\mbox{ if } \{f,g\}=0 \mbox{ for all } g \mbox{, then } f=const$.
This canonical Poisson bracket equips the phase space with a symplectic structure
\cite{Marsden}.  The Hamiltonian dynamics is then determined by defining the proper
Hamiltonian function $\calH$.  The evolution equation for any phase space function
$f(q,p)$ reads then: $\frac{df}{dt} = \frac{\pa f}{\pa t} + \{f, \calH \}$.

  In applications one often encounters a situation when the phase space dynamics
is subject to certain {\em external} restricting conditions on the phase space
variables called constraints. Often the constraints can be written in terms
of some phase space functions $\phi_i(q,p)=0$, and we will restrict our analysis
to these cases only. The Hamiltonian formalism for such constrained systems requires
modifications. These modifications have been first suggested by Dirac \cite{Dirac1},
and a brief account of the Dirac theory follows.

  Let $\phi_i$ (with $i= 1,\ldots,L$) denote all constraints for our Hamiltonian
system.  Those constraints can be divided into two classes by analyzing the
$L \times L$ skew-symmetric matrix of their mutual Poisson brackets
$A_{i\,j} = \{\phi_i,\phi_j\}$.
  Since $A$ is skew-symmetric, its rank $K$ must be even. We assume that after
relabeling of the $\phi_i$ and/or redefining the constraints by taking their
linear combinations (known as the Dirac separating constraints algorithm), the
top left $K \times K$ submatrix of $A$, which we denote by $W$, is regular.
The constraint functions $\phi_{K+1},\ldots,\phi_L$ are then called first class
constraints, and are associated with local gauge symmetries \cite{Dirac1},
while $\phi_1,\ldots,\phi_K$ are called second-class.  In this work we will
consider second-class constraints only, and for them we can introduce the
{\it Dirac bracket} (DB)\cite{Dirac1}, of two phase space functions $f,g$:
\beqn \label{DB00}
   \{f,g\}_D &=& \{f,g \} -
                 \sum_{i,j=1}^{K} \{f,\phi_i\} (W^{-1})_{i j} \{\phi_j,g\} \,.
\eeqn
  In the modern language of symplectic geometry, constrained Hamiltonian dynamics
can be represented by a triplet $(M,N,\omega)$ where $(M,\omega)$ is a symplectic
manifold, namely Phase space, and $N$ is a constraint submanifold of $M$.  The DB
(\ref{DB00}) is the Poisson bracket on a symplectic submanifold $N' \subset N$,
called second-class constraint manifold \cite{Marsden,Flato,Sniatycki,Bhaskara}.

  Symplectic structure requires even dimensional manifolds and non-degenerate
Poisson structure. Both these assumptions seem too restrictive and not always
applicable. With the appearance of non-canonical Poisson structure (PS) in
rigid body dynamics, theory of magnetism, infinite dimensional PS in
magneto-hydrodynamics, etc. and issues of geometric quantization, systematic studies
of the general {\it Poisson bracket} (PB) which is a Lie bracket satisfying the 
Leibniz identity,
has become important.

  The fundamental geometric object in the description of any generalized
Hamiltonian dynamics is a Poisson manifold.  Geometrically, Poisson manifold 
is a manifold endowed with a bivector field $\pi$ satisfying $[\pi, \pi] = 0$,
where $[\cdot, \cdot]$ denotes the Schouten bracket\cite{Schouten} on
multivector fields.
  Algebraically, $M$ is a Poisson manifold if there is a Poisson bracket on the
space of smooth functions defined on $M$.  The Poisson bracket $\{\cdot, \cdot\}$
and the bivector field $\pi$ determine each other \cite{Bhaskara,Vaisman} by the
formula $\{f, g\} = \pi(df, dg)$. Both the geometric and algebraic characterization
of Poisson manifolds are used in the literature.

  In the analysis of the constrained systems dynamics it is of predominant importance
to formulate it as a usual Poisson structure on a submanifold of a non-constrained
system's Poisson manifold.  The conditions under which the Poisson structure on a
submanifold is achievable was investigated in \cite{Xu,Crainic} and the geometric
derivation of the DB formula (\ref{DB00}) via a procedure called geometric reduction
of Poisson tensor was known \cite{Marciniak}.

  In many of the important physical applications the systems described are not
purely Hamiltonian but also dissipative.  The description of such combined
dissipative-hamiltonian dynamics can be formulated in various ways, however one
of them seems to be particularly elegant and allows to incorporate in it many
methods developed in purely symplectic dynamics.
  This method was introduced first in the phase transformation kinetics in
\cite{Langer-Turski} and then independently in \cite{Enz-Turski, Morrison} and
called metriplectic.
  The main point in metriplectic formulation \cite{Morrison} is that a mixed bracket
obtained by adding a symmetric bracket to the Poisson bracket can successfully be
used for description of dissipative systems.

  In the metriplectic framework, the underlying structure of a dissipative system
consists of a Poisson and a symmetric bracket \cite{Morrison}, and the obvious
generalization of this construction for {\it constrained dissipative system} (CDS) 
must consist of two DB \cite{SNT3}: the usual skew-symmetric DB and the symmetric DB,
which describe the Hamiltonian and dissipative part respectively.
  In \cite{SNT3} we have assumed that CDS be geometrically represented by a triplet 
$(M,N,\omega - g)$, here $N$ is a submanifold of the symplectic manifold $(M, \omega)$ 
and $g$ is a covariant semimetric tensor. Generalized result can be easily obtained by
replacing the symplectic $2$-form $\omega$ by a contravariant Poisson tensor $\pi$, and
the covariant metric $(0,2)$ tensor $g$ by a contravariant (semi/pseudo)-metric $(2,0)$ 
tensor $G$.

  The aim of the article is to give a formal (algebraic) proof of the recursiveness of
symmetric and skew-symmetric DB. For the latter, this property probably has been
known for years in practical calculation, but none algebraic proof seems to be
available in the literature.
  The proof given in this paper is, to the best of our knowledge, the first one.

The paper is organized as follows.
  Section $2$ resumes a construction leading to the DB-like formula in the 
general case and conditions of submanifold possessing Poisson structure 
in the form of the DB.
  Section $3$ presents rigorous proof for the recursiveness of symmetric and
skew-symmetric DB.
  Section $4$ illustrates the constrained metriplectic formalism on two examples,
using the computer algebra package Mathematica.
  Appendix A shows that symbolic/analytical difficulties appeared in the Dirac
approach are unavoidable and that they also appear in the Lagrangian approach.
  Appendix B contains Dirac and LMM description for $N$-pendulum, which serves
as our numerical case study.
  Appendix C contains some techniques for analytical inversion of 
symmetric tridiagonal matrices, which we worked out in 2004.

  In this article, we denote a symmetric, skew-symmetric and general bracket by
$<\cdot,\cdot>$, $\{\cdot,\cdot\}$ and $\eta(\cdot,\cdot)$ respectively.

\sect{Geometric interpretation on Dirac-like brackets}

  We begin by showing how an arbitrary $K$-tensor defined on a manifold $M$ can
be reduced in the (almost) Dirac sense to any submanifold of $M$, regardless of
this tensor degeneracy.

  Conventionally we will denote the dual spaces to $E$ , $F$, etc. and similar
dual map to $f$, etc. by superscript asterisk, e.g $E^*$, $F^*$ and $f^*$,
the annihilator\footnote{The annihilator of $F \subset E$ is
$F^0 = \{\phi \in E^* \mbox{ such that } \phi(F)=0 \}$} by superscript zero,
e.g $F^0$ and $V^0$.
  Furthermore, denoting annihilation between elements of $E$ and $E^*$ by
$(\cdot \,|\, \cdot)$, each bivector $\pi \in \wedge^2 E$ defines the map
$\pi^{\sharp}: E^* \to E$ by $(\pi^{\sharp}(\zeta) \,|\, \eta) = \pi(\zeta,\eta)$
for $\zeta,\, \eta \in E^*$.
  The term almost Poisson structure means that this structure is bilinear and
skew-symmetric, but does not necessarily satisfy the Jacobi identity.

Let $E$ be a linear space, $F$ be its linear subspace and let $E$ be a direct
sum $E = F \oplus V$.   This direct sum determines uniquely the projection
$p: E \to F$ and induces a splitting in its dual space 
$E^* = F^* \oplus V^*$ with $F^* = V^0$ and $V^* = F^0$.  The direct 
decomposition on $E^*$ determines uniquely the map $p^*: F^* \to E^*$ 
which is the dual map of $p$. 

\begin{defi}
{\em  Each multilinear map $K: (E^*)^k \to R$ induces multilinear map 
      $K_F: (F^*)^k \to R$ by
      \beqn
          K_F (\alpha_1,\cdots,\alpha_k) &=& 
          K ( p^* (\alpha_1),\cdots,p^* (\alpha_k) ) \,. 
      \eeqn  
      We will call $K_F$ an almost Dirac reduction of $K$ on $F$ with respect 
      to the direct sum $E=F \oplus V$ (or with respect to the projection $p$).  
}
\end{defi}

If $K$ is symmetric or skew-symmetric then $K_F$ has the same properties, but 
$K_F$ may not inherit other algebraic properties of $K$.  In particular, 
if $K$ is non-negative, i.e. 
$K(\alpha,\alpha,\cdots,\alpha) \geq 0, \; \forall \alpha \in E^*$, then $K_F$ 
also is non-negative, but if $K = \pi$ is Poissonian,  the bracket defined 
by $\pi_F$ may not satisfy the Jacobi identity. Thus, in general $\pi_F$ is only 
{\it almost} Poissonian.
The sufficient condition for $\pi_F$ to be Poissonian is:
\begin{prop} \label{P-Dirac subspace}
Suppose that $\pi$ is a Poisson tensor in $E$ such that
$F ~\cap ~\pi^{\sharp} (F^0) = \{ 0 \}$.  Then $\pi_F$ is a Poisson tensor.
\end{prop}
{\bf Proof.}  Since $F \cap \pi^{\sharp} (F^0) = \{ 0 \}$, 
one has $\pi^{\sharp}(\zeta) \in V$ for every $\zeta \in F^0$. 
Hence $\pi(\zeta,\eta)= ( \pi^{\sharp}(\zeta) \,|\, \eta ) = 0$ for every 
$\zeta \in F^0, \eta \in V^0$. 
This orthogonality condition implies that $\pi \in \wedge^2 E$ decomposes as
$\pi = \pi_F + \pi_V$ where $\pi_F \in \wedge^2 F$ and $\pi_V \in \wedge^2 V$.
The identity $[\pi,\pi]=0$ implies that 
$[\pi_F,\pi_F]=[\pi_F+\pi,\pi_F-\pi]=-[2\pi_F+\pi_V,\pi_V]$.  Because of
$[\pi_F,\pi_F] \in \wedge^3 F \subset \wedge^2 E \wedge F$ and 
$-[2\pi_F+\pi_V,\pi_V] \in \wedge^2 E \wedge V$, they must both be zeros. 
Therefore $[\pi_F,\pi_F]=0$ which means that $\pi_F$ is Poissonian. 

Proposition (\ref{P-Dirac subspace}) leads to the following concept of 
Dirac subspace:   A linear subspace $F$ of a Poisson space $(E,\pi)$ 
is called a {\em Dirac subspace} if $F \cap  \pi^{\sharp} (F^0) = \{ 0 \}$.  
Furthermore, since any projection $p: E \to F$ defines a split $E=F \oplus V_p$ 
where $V_p=(Id-p) F$ and 
the condition $F \cap  \pi^{\sharp} (F^0) = \{ 0 \}$ is equivalent to 
$\pi^{\sharp}(F^0) \subset V_p$,  this condition suggests to introduce a 
concept of Dirac projection:  A linear map $p: E \to F$ is called a 
{\it Dirac projection} if $p(\pi^\sharp(F^0)) = 0$.

We now consider the non-linear case. 
Let us consider a smooth finite dimensional manifold $M$, a submanifold 
$N \subset M$ and a regular distribution $\calV$ on $M$ (that is a smooth 
family of the subspaces of the tangent spaces, $\calV_{x} \subset T_{x}M$) 
such that $T_{x} M = T_{x} N \oplus \calV_x$ for every $x$ in $N$. Thus
$\calV$ is {\it complementary} of $T N$ in $T M$.  
  
For any $k$ one-forms $\alpha_1,\cdots,\alpha_k$, the reduction of 
$(k,0)$-tensor field $K$ on $N$ is defined by:  
\beqn
   K_{N}\left( \alpha_1,\cdots,\alpha_k\right)  = 
   K( p^* (\alpha_1),\cdots,p^* (\alpha_k) ) \,.
\eeqn
We call the tensor field $K_{N}$ the almost Dirac reduction of $K$ with respect 
to the submanifold $N$ and the direct decomposition $T_{N} M = T N \oplus \calV$.  

Applying proposition (\ref{P-Dirac subspace}) one gets the following
\begin{prop} \label{P-Dirac submanifold}
Let $N$ be a submanifold of a Poisson manifold $(M,\pi)$.  Suppose that
$T N \cap \pi^{\sharp} (T N^0) = \{0\}$ and $\pi_N$ is smooth. Then,
$\pi_N$ is a Poisson tensor on $N$. 
\end{prop}

Thus we obtain a sufficient condition for constructing Poisson structure 
on a submanifold. Furthermore, proposition (\ref{P-Dirac submanifold}) 
leads to the following concept of Dirac submanifold.
\begin{defi} \label{dirac_sub_defi} \cite{Crainic}
A submanifold $N$ of the Poisson manifold $(M,\pi)$ is called a {\em Dirac 
submanifold} if $T N \cap  \pi^{\sharp} (T N^0) = \{ 0 \}$ and induced 
tensor $\pi_N$ is smooth.
\end{defi}
Note that the concept of Dirac submanifold (def. \ref{dirac_sub_defi}) is less
restrictive than the one introduced by Xu \cite{Xu} and the other mentioned
therein:
A submanifold $N$ is -called by Xu- a Dirac submanifold of the Poisson manifold
$(M,\Pi)$ if there exists a bundle $\calV$ such that $T_N M = T N \oplus \calV$
and $\calV$ is a coisotropic submanifold of $T M$.


  For applications, the most important case of this geometric procedure is
when $K=\pi \pm  G$, where $\pi, G$ are Poisson and pseudo/semi-metric tensor, 
respectively.

%
%
\sect{Algebraic formulas for computing Dirac brackets}
%
%
\subsect{Pfaffians and the Tanner's identities}

For any function of two arguments $F$ defined on the set of generators of 
the commutative algebra $\calA$, we introduce the notation
\beqn \label{pfaff-notes}
    F[x_1 \cdots x_n, y_1 \cdots y_n] = \det ( F[x_i,y_j] ) = 
    \left|\begin{array}{cccc} 
                       F[x_1,y_1] & \cdots & F[x_1,y_n] \\
                           \vdots &        & \vdots \\
                       F[x_n,y_1] & \cdots & F[x_n,y_n] 
           \end{array} 
    \right|\,.
\eeqn
We will use the following identities:
\beqn    
\label{tanner-2}
    F[\alpha,\beta] \, F[\alpha x z, \beta y t] &=& 
       F[\alpha x, \beta y] \, F[\alpha z, \beta t] - 
       F[\alpha x, \beta t] \, F[ \alpha z, \beta y]   \\
\label{tanner-3}
    F[\alpha,\beta] \, F[\alpha x u v, \beta y s t] &=& 
       F[\alpha x, \beta y] \, F[\alpha u v, \beta s t] - 
       F[\alpha x, \beta s] \, F[ \alpha u v, \beta y t] \nonumber \\
       &+& F[\alpha x,\beta t] \, F[ \alpha u v,\beta y s] \,,
\eeqn
which are a special case of the Tanner identity \cite{Tanner,Knuth}; and they also 
are known as theorems on bordered determinants \cite{Vein-Dale}, pages 46-50.   
Assuming $F[u,v]=\eta(u,v)$ for $u, v$ from a commutative algebra with the bracket 
$\eta$, we have
\beqn
  F[\phi_1 \cdots \phi_N,\xi_1 \cdots \xi_N] &=&
  \left|\begin{array}{cccc} 
        \eta(\phi_1,\xi_1) & \cdots & \eta(\phi_1,\xi_N) \\
                    \vdots &        & \vdots \\
        \eta(\phi_N,\xi_1) & \cdots & \eta(\phi_N,\xi_N) \\
        \end{array} 
  \right|.
\eeqn
%
%
\subsect{Determinant and recursive formulas}

Let $(\calF,\cdot)$ be a commutative algebra with the bracket 
$\eta: \calF \times \calF \to \calF$ 
and $\{\phi_i\}_{i=1}^{n}$ be a set of elements from $\calF$.  
Suppose the square matrix $W = (W_{ij})$ with $W_{ij}=\eta(\phi_i,\phi_j)$ 
is invertible, and let us denote its inverse matrix by $C=[C_{ij}]$.  
The original DB formula follows:
\beqn \label{Dirac_formula_gene1}
    \eta_D(f,g) &=& \eta(f,g) -\sum_{i,j=1}^{n} 
    \eta(f,\phi_i) \, C_{ij} \, \eta(\phi_j,g) \,, ~\forall f,g \in \calF \,.
\eeqn
The new bracket (\ref{Dirac_formula_gene1}) is bilinear and it inherits 
algebraic properties from the original bracket $\eta$.  It is easy to 
check that $\forall f\in \calF$, $\eta_D(\phi_i,f) = 0$, which means that 
all elements $\phi_i$ are in the algebra center (called Casimir's elements) 
of the algebra $(\calF,\eta_D)$.
For skew-symmetric algebras the number of fixed elements $\phi_j$ must be 
even, because the skew-symmetric matrix $W$ with odd rank always is singular.  
Indeed, denoting $\det W$ by $|W|$, for skew-symmetric matrix $W$ we have 
$|W| = |W^T| =(-1)^n |W|$.   
%
%


Let $A=(a_{ij})$ be a matrix, then the matrix obtained from $A$ after 
deleting $i-$th row and $j-$th column will be denoted by $A^{(i,j)}$.  
Recall the Laplace expansion formula which states that
$\det A = |A| = \sum_{j} (-1)^{i+j} a_{ij} |A^{(i,j)}|$ for any square 
matrix $A$.
Now we can easily prove the following {\em determinant formula} for the DB.
\begin{prop} \cite{SNT3}
\label{prop_S1}
{\em
Supposing the matrix $W(\phi_1,\ldots,\phi_n)$ is invertible, 
the following identity holds 
\beqn \label{SN-formula1-matrix}
      \forall f, g  \in \calF \,: ~~
      \eta_D(f,g) &=& 
           \frac{\left|
                 \begin{array}{cccc}
                     \eta(\phi_1,\phi_1) & \cdots & \eta(\phi_1,\phi_n) & \eta(\phi_1,g)\\
                     \vdots              & \vdots & \vdots              & \vdots \\       
                     \eta(\phi_n,\phi_1) & \cdots & \eta(\phi_n,\phi_n) & \eta(\phi_n,g)\\
                     \eta(f,\phi_1) & \cdots & \eta(f,\phi_n) & \eta(f,g)    
                 \end{array} \right| }{
                 \left|
                 \begin{array}{ccc}
                     \eta(\phi_1,\phi_1) & \cdots & \eta(\phi_1,\phi_n) \\
                     \vdots              & \vdots & \vdots              \\       
                     \eta(\phi_n,\phi_1) & \cdots & \eta(\phi_n,\phi_n) 
                 \end{array} \right| } \,.    
\eeqn
Rewriting (\ref{SN-formula1-matrix}) in the notation (\ref{pfaff-notes}) we get
\beqn \label{SN-formula1}
   \forall f,g \in \calF \,: ~~
   \eta_D(f,g) &=& \frac{ |W_{f,g}| }{ |W| } \,, 
\eeqn
where $|W| = F[\phi_1 \cdots \phi_n , \phi_1 \cdots \phi_n]$ and
$|W_{f,g}| = F[\phi_1 \cdots \phi_n f, \phi_1 \cdots \phi_n g]$.
}
\end{prop}
{\bf Proof.}  
Apply twice the Laplace formula to the last column and row of the matrix 
$W_{f,g}$. \\
{\bf A. Symmetric case:}\\
Now let $(\calF,\cdot)$ be a commutative algebra with the bracket $<\cdot,\cdot>$ 
and $\{\phi_j\}_{j=1}^n$, 
be a set of elements from $\calF$.  We define inductively a family of brackets
\beqn \label{recursive00}
    <f,g>^{(0)} &=& <f,g> \,, \nonumber\\   
    < f,g >^{(k+1)} &=& < f,g >^{(k)} -
    \frac{ < f,\phi_{k+1} >^{(k)} < \phi_{k+1},g >^{(k)} }
         { < \phi_{k+1},\phi_{k+1} >^{(k)} } \,.
\eeqn 

Denote the Dirac bracket determined by $k$ constraints $\phi_a$ with 
$a=1,\cdots,k$, by $< f,g >_D^{(k)}$, thus  
\beqn
   < f,g >_D^{(k)} = 
   < f,g > -\sum_{a,b=1}^k < f,\phi_a > \, C_{ab}^{(k)} \, < \phi_b,g > \,, 
\eeqn
where $C^{(k)}$ is the inverse matrix of $k \times k$ matrix 
$W^{(k)}=\left[ \begin{array}{ccc}
                     \eta(\phi_1,\phi_1) & \cdots & \eta(\phi_1,\phi_k) \\
                     \vdots              & \vdots & \vdots              \\
                     \eta(\phi_k,\phi_1) & \cdots & \eta(\phi_k,\phi_k) 
                 \end{array} \right]$.

We prove the following theorem
\begin{theo}[Recursive general brackets]
\label{theo_S1}
{\em

Assume that 
the family of brackets (\ref{recursive00}) is well-defined.   
Then  $\forall f,g \in \calF$ and $1 \leq m \leq n$:
\beqn \label{recursive01}
    < f,g >^{(m)} &=& < f,g >_D^{(m)} \,. 
\eeqn
}
\end{theo}

{\bf Proof.} We prove the formula (\ref{recursive01}) by induction with 
$m$.  For $m=1$, (\ref{recursive01}) is obviously true. 
Suppose that it is true for $m=k$, thus 
\beqn \label{induct-assum}
   \forall f, g:~~ < f,g >^{(k)} = < f,g >_D^{(k)},
\eeqn 
we shall prove that it remains true for $m=k+1$.  The proof is based on 
the Tanner identity (\ref{tanner-2}) and the proposition {\bf \ref{prop_S1}}.  

First, let $\alpha=\phi_1\phi_2\cdots\phi_{k}$, using formula (\ref{SN-formula1}) 
in the proposition {\bf \ref{prop_S1}} we have
\beqn \label{eq38}
  < f,g >_D^{(k+1)}= \frac{F[\alpha\, \phi_{k+1}f,\alpha \,\phi_{k+1}g]}
                      {F[\alpha\, \phi_{k+1},\alpha \,\phi_{k+1}]} \,.
\eeqn
Multiplying r.h.s. of (\ref{eq38}) by $1=\frac{F[\alpha,\alpha]}{F[\alpha,\alpha]}$ 
and using (\ref{tanner-2}) we get
\beqn \label{eq39}
     < f,g >_D^{(k+1)} &=& \frac{F[\alpha f,\alpha g]}{F[\alpha,\alpha]}-
     \frac{F[\alpha f,\alpha \,\phi_{k+1}]\, F[\alpha \, \phi_{k+1},\alpha g]}
          {F[\alpha,\alpha] \, F[\alpha \,\phi_{k+1},\alpha \,\phi_{k+1}]} 
\eeqn
Using formula (\ref{SN-formula1}) again, we show that: the first term in the 
r.h.s. of eq. (\ref{eq39}) is equal $< f,g >_D^{(k)}$ and also equal 
$< f,g >^{(k)}$ by induction assumption (\ref{induct-assum}).  Applying similar 
argument for the second term in the r.h.s. of eq. (\ref{eq39}), we obtain
\beqnn 
   \frac{F[\alpha f,\alpha \,\phi_{k+1}]} {F[\alpha,\alpha]}
   = <f,\phi_{k+1}>^{(k)} \,, ~
   \frac{F[\alpha \, \phi_{k+1},\alpha g]}{F[\alpha,\alpha]}
   =<\phi_{k+1},g>^{(k)}  ~ \mbox{ and } \\
   \frac{F[\alpha \,\phi_{k+1},\alpha \,\phi_{k+1}]}{F[\alpha,\alpha]} 
   = <\phi_{k+1},\phi_{k+1}>^{(k)}.
\eeqnn
In summary, the r.h.s. of eq. (\ref{eq39}) is equal  
\beqn
   < f,g >^{(k)}-\frac{ <f,\phi_{k+1}>^{(k)} <\phi_{k+1},g>^{(k)}}
                      {<\phi_{k+1},\phi_{k+1}>^{(k)}} \,.
\eeqn
It implies that r.h.s. of eq. (\ref{eq39}) is equal $< f,g >^{(k+1)}$ 
which ends the proof.
$\spadesuit$ \\

To apply theorem {\bf \ref{theo_S1}} we need an existence of the family of 
brackets (\ref{recursive00}). This condition requires the invertibility of 
$<\phi_{i+1},\phi_{i+1}>^{(i)}$ for all $i$ with $1 \leq i \leq n$, and 
therefore it is equivalent to the regularity (or non-degeneracy) of 
all main minors of $W$. 
This condition may seem to be too restrictive, however by making new constraints
from linear combinations of old constraints, we can go beyond this restriction
The following simple example illustrates the procedure.
\begin{exam}
{\em
Let $x=(x_1,x_2,\cdots,x_n) \in R^n$, 
\beqnn
   <x_1,x_1>=<x_2,x_2>=0, ~ <x_1,x_2>=<x_2,x_1>=a(x),
\eeqnn
other brackets are whatever, and the constraints are
$\phi_1 = x_1 = 0$, $\phi_2 = x_2 = 0$.

In the standard approach, after calculating the constraint matrix 
$W=a(x) \left[\begin{array}{cc}
                      0 & 1 \\ 1 & 0  
              \end{array} 
        \right]$, and its inverse, we easily get the Dirac bracket
\[   <f,g>_D = <f,g> - \frac{1}{a(x)}(<f,x_1><x_2,g> + <f,x_2><x_1,g>) \,.\]
In this case, direct recursive scheme is inapplicable because of
\[    <\phi_1,\phi_1>~ = ~0~ = ~<\phi_2,\phi_2> \,,\]
but by introducing new (equivalent) constraints 
$u_1=x_1+x_2=0$ and $u_2=x_1-x_2=0$,
the recursive scheme may apply as below.

In the first step, we have
\beqnn
    <f,g>^{(1)}&=&<f,g>-\frac{<f,u_1><u_1,g>}{<u_1,u_1>} \,.
\eeqnn
Since $<u_1,u_2>=0$ we get $<f,u_2>^{(1)}=<f,u_2>$, $<u_2,g>^{(1)}=<u_2,g>$ and 
$<u_2,u_2>^{(1)}=<u_2,u_2>$. Hence,
\beqnn
  <f,g>^{(2)}&=&<f,g>^{(1)}-\frac{<f,u_2>^{(1)}<u_2,g>^{(1)}}{<u_2,u_2>^{(1)}}
                \nonumber\\
  &=&<f,g>-\frac{<f,u_1><u_1,g>}{<u_1,u_1>}-\frac{<f,u_2><u_2,g>}{<u_2,u_2>}.
\eeqnn
Finally, express it in terms of the original constraints
\beqnn
 <f,g>^{(2)} &=& <f,g> - \frac{1}{a(x)}(<f,x_1><x_2,g> + <f,x_2><x_1,g>)\,.
\eeqnn  
}
\end{exam}

We can use theorem \ref{theo_S1} to prove that symmetric DB inherits
non-negativity from a semimetric bracket.  Precisely,
\begin{prop}
{\em
Suppose $\calF$ be an algebra of real functions with semimetric bracket
$<\cdot,\cdot>$, i.e. $<f,f>$ is a non-negative function for every function
$f \in \calF$.  Let $\{\phi_k \}_{k=1}^{n}$ be a set of elements from
$\calF$ such that $W(\phi_1,\ldots,\phi_{n})$ is invertible.
Then the Dirac bracket $<\cdot,\cdot>_D$ with respect to $\{\phi_k\}_{k=1}^{n}$,
is semimetric.
}
\end{prop}
{\bf Proof.}
Since the recursion property of symmetric DB in theorem \ref{theo_S1},
it is enough to prove $<f,f>^{(1)}$ is a non-negative function.
Indeed, for every real number $\lambda$, one has
\beqnn
    0 \leq  \; <f - \lambda \phi_1,f - \lambda \phi_1> = 
               <f,f> - 2 \lambda <f,\phi_1> + \lambda^2 <\phi_1,\phi_1> \,,
\eeqnn
which implies that the discriminant
$\triangle = [<f,\phi_1>]^2-<f,f> <\phi_1,\phi_1> \leq 0$.
Thus,
\beqnn
	<f,f>^{(1)} = <f,f> - \frac{<f,\phi_1>^2}{<\phi_1,\phi_1>} \geq 0 \,.
\eeqnn
{\bf B. Skew-symmetric case:}\\
Now let $(\calF,\cdot)$ be a commutative algebra with a skew-symmetric 
bracket $\{\cdot,\cdot\}$ and $\{\phi_k\}_{k=1}^{2n}$, 
be a set of elements from $\calF$.   We define inductively a family of brackets
{\small
\beqn \label{recursive-anti00}
    \{f,g\}^{(0)} &=& \{f,g\}, \nonumber \nopagebreak\\
    \{f,g\}^{(k+1)} &=& \{f,g\}^{(k)} -
    \frac{ \{f,\phi_{2k+2}\}^{(k)} \{\phi_{2k+1},g\}^{(k)}
          -\{f,\phi_{2k+1}\}^{(k)} \{\phi_{2k+2},g\}^{(k)} }
         {\{\phi_{2k+1},\phi_{2k+2} \}^{(k)}}.  \nonumber\\  
\eeqn 
}
We prove that (\ref{recursive-anti00}) are identical with the Dirac brackets.
\begin{theo}[Recursive skew-symmetric brackets]\label{theo_S2}\
{\em
Suppose that the family of bracket recursively defined by 
(\ref{recursive-anti00}) is well-defined. Then $\forall f,g \in \calF$ 
and $1 \leq m \leq n$:
\beqn \label{recursive-anti01}
    \{f,g\}^{(m)} &=& \{f,g\}^{(2m)}_D \,,
\eeqn
where r.h.s. is the Dirac bracket with respect to $2m$ constraints
\[
    \{f,g\}^{(2m)}_D = \{f,g\} -\sum_{a,b=1}^{2m} 
    		       \{f,\phi_a \} \, C_{ab}^{(2m)} \, \{\phi_b,g\} \,.
\]
In the above $C^{(2m)}$ in the inverse of the $2m \times 2m$ matrix $W^{(2m)}$
\beqnn
    W^{(2m)}=\left[\begin{array}{ccc}
                      \{\phi_1,\phi_1\} & \cdots & \{\phi_1,\phi_{2m} \} \\
                      \vdots              & \vdots & \vdots              \\       
                      \{\phi_{2m},\phi_1\} & \cdots & \{\phi_{2m},\phi_{2m}\} 
                   \end{array} \right].
\eeqnn
}
\end{theo}
{\bf Proof.}  We prove this theorem by induction with $m$.

It is true for $m=1$ and suppose that $\{f,g\}^{(k)}=\{f,g\}_D^{(2k)}$ for some
$k \geq 1$, we shall prove that $\{f,g\}^{(k+1)}=\{f,g\}_D^{(2k+2)}$.
Let denote $\alpha=\phi_1 \cdots \phi_{2k}$, because of (\ref{SN-formula1}) 
in the proposition {\bf \ref{prop_S1}} we have:
\beqn \label{eq312}
  \{f,g\}^{(2k+2)}_D = 
  \frac{ F[\alpha \phi_{2k+1} \phi_{2k+2} f, \alpha \phi_{2k+1} \phi_{2k+2} g]}
       { F[\alpha \phi_{2k+1} \phi_{2k+2}, \alpha \phi_{2k+1} \phi_{2k+2}]} .
\eeqn
Multiplying r.h.s. of (\ref{eq312}) by $1=\frac{F[\alpha,\alpha]}{F[\alpha,\alpha]}$, 
using the Tanner identities (\ref{tanner-2}), (\ref{tanner-3}) and knowing 
determinant of a skew-symmetric matrix of odd size to be zero, 
$F[\alpha \phi_{2k+1},\alpha \phi_{2k+1}]=0$, we get the r.h.s of (\ref{eq312})
{\small
\beqnn
      \frac{ F[\alpha\phi_{2k+1},\alpha g]\, 
             F[\alpha\phi_{2k+2}f,\alpha\phi_{2k+1}\phi_{2k+2}]
            -F[\alpha\phi_{2k+1},\alpha \phi_{2k+2}]\, 
             F[\alpha\phi_{2k+2}f,\alpha\phi_{2k+1} g]}
           {-F[\alpha\phi_{2k+1},\alpha \phi_{2k+2}]\, 
             F[\alpha\phi_{2k+2},\alpha\phi_{2k+1}]} .
\eeqnn
}
Again, multiplying by $1=\frac{F[\alpha,\alpha]}{F[\alpha,\alpha]}$, using 
the Tanner identities (\ref{tanner-2}), the vanishing determinant
of a skew-symmetric matrix of odd size, i.e.
$F[\alpha \phi_{2k+2},\alpha \phi_{2k+2}]=0$, and the recursive assumption
$\{u,v\}^{(k)}=\{u,v\}_D^{(2k)}$ we obtain:
{\small
\beqnn
    \{f,g\}^{(2k+2)}_D &=& \frac{F[\alpha f,\alpha g]}{F[\alpha,\alpha]}+
    \frac{F[\alpha f,\alpha \phi_{2k+1}] F[\alpha \phi_{2k+2},\alpha g]-
          F[\alpha f,\alpha \phi_{2k+2}] F[\alpha \phi_{2k+1},\alpha g]}
         {F[\alpha,\alpha]F[\alpha\phi_{2k+1},\alpha\phi_{2k+2}]} \nonumber \\
    &=&  \{f,g\}^{(2k)}_D + \frac{ \{f,\phi_{2k+1}\}^{(2k)}_D \{\phi_{2k+2},g\}^{(2k)}_D
                                   -\{f,\phi_{2k+2}\}^{(2k)}_D \{\phi_{2k+1},g\}^{(2k)}_D }
                                 { \{\phi_{2k+1},\phi_{2k+2}\}^{(2k)}_D } \nonumber\\
    &=& \{f,g\}^{(k)}+ \frac{ \{f,\phi_{2k+1}\}^{(k)}\{\phi_{2k+2},g\}^{(k)}
                           -\{f,\phi_{2k+2}\}^{(k)}\{\phi_{2k+1},g\}^{(k)}}
                          { \{\phi_{2k+1},\phi_{2k+2}\}^{(k)} } \,.
\eeqnn
}
It implies that r.h.s. of eq. (\ref{eq312}) is equal $\{f,g\}^{(k+1)}$ 
which ends the proof.
$\spadesuit$\\

Theorems {\bf \ref{theo_S1}} and {\bf \ref{theo_S2}} are main results of 
this article.

One may use theorem \ref{theo_S2} in proving Jacobi identity and some 
other algebraic properties for Dirac bracket. For example, one can prove 
the following
\begin{prop}
{\em
Suppose $(\calF,\cdot,\{\cdot,\cdot\})$ be skew-symmetric algebra and 
$\{\phi_k, ~ k =1,\ldots,2n \}$ be a set of elements from $\calF$ such that
$W(\phi_1,\ldots,\phi_{2n})$ is invertible. Then $\forall f, g \in \calF$:
\beqn     \label{quad-dirac0}
     \{f,g\}_D^2 = \frac{| W(\phi_1,\ldots,\phi_{2n},f,g) |}
                          {| W (\phi_1,\ldots,\phi_{2n}) |}
                 = \frac{F[\phi_1 \cdots \phi_{2n} f g, \phi_1 \cdots \phi_{2n} f g]}
                        {F[\phi_1 \cdots \phi_{2n},\phi_1 \cdots \phi_{2n}]} \,.
\eeqn
}
\end{prop}
{\bf Proof.}  Let $\alpha=\phi_1\phi_2\cdots\phi_{2n}$.  From the identity
(\ref{tanner-2}) we have
{\small
\beqn \label{quad-dirac1}
    F[\alpha,\alpha] \, F[\alpha f g, \alpha f g]
    =      F[\alpha f,\alpha f] F[\alpha g, \alpha g] -
           F[\alpha f,\alpha g] F[\alpha g, \alpha f]
    = (F[\alpha f,\alpha g])^2. \nonumber\\
\eeqn 
}
Dividing both sides of (\ref{quad-dirac1}) by $(F[\alpha,\alpha])^2$
(i.e. $| W (\phi_1,\ldots,\phi_{2n}) |^2$ ) we obtain
$\frac{F[\alpha fg,\alpha fg]}{F[\alpha,\alpha]}=\{f,g\}_D^2$.

\subsect{Jacobi identity}
In \cite{Dirac1}, Dirac was struggling to prove the Jacobi identity for 
his bracket formula.  He wrote: "I think  there ought 
to be some neat way of proving it, but I haven't been able to find it". 
The Proposition \ref{jacobi-dirac01} below contains what we believe is 
just that kind of a proof.

\begin{prop} \label{jacobi-dirac01}

{\em  Let $(\calF,\cdot)$ be a commutative algebra with Lie or 
      Poisson bracket $\{\cdot,\cdot\}$. Suppose $\{\phi_k, ~ k=1,\ldots,2n \}$ 
      be a set of elements from $\calF$ such that $(\{\phi_i,\phi_j\})$ is 
      invertible.
      Then $\{\cdot,\cdot\}_D$ with respect to $\{\phi_k\}_{k=1}^{2n}$
      is a Lie or Poisson bracket, respectively. 
}
\end{prop}
{\bf Proof.}~  Only the Jacobi identity is difficult to verify.  Using the 
theorem \ref{theo_S2} and the induction principle, it is enough to show 
that $\{\cdot,\cdot\}^{(1)}$ satisfies the Jacobi identity.
  In order to check the Jacobi identity for $\{\cdot,\cdot\}^{(1)}$, it is 
convenient to introduce the following symbols:  
$A_i=\{f,\phi_i\}\,,B_i=\{g,\phi_i\}\,,C_i=\{h,\phi_i\}$ with $i=1,2$ and
$\phi_{12}=\{\phi_1,\phi_2\}$.
Since the Jacobi identity holds for $\{\cdot,\cdot\}$ all the following 
sums vanish
{\small
\beqn \label{jacobi-help1}
     I_i=\{A_i,g\} + \{f,B_i\} + \{\phi_i,\{f,g\}\}, ~
     J_i=\{A_i,h\} + \{f,C_i\} + \{\phi_i,\{f,h\}\}, \nonumber\\
     K_i=\{C_i,g\} + \{h,B_i\} + \{\phi_i,\{h,g\}\}, ~
     D=\{\phi_2,A_1\}+\{A_2,\phi_1\}+\{f,\phi_{12}\}, \nonumber\\
     E=\{\phi_2,B_1\}+\{B_2,\phi_1\}+\{g,\phi_{12}\}, ~
     F=\{\phi_2,C_1\}+\{C_2,\phi_1\}+\{h,\phi_{12}\}. \nonumber\\
\eeqn
}
Full expansion of $Jacobi=\{f,\{g,h\}_D\}_D + \{g,\{h,f\}_D\}_D + \{h,\{f,g\}_D\}_D$ 
produces 39 non-vanishing terms that can be grouped in a polynomial of the 
variable $z=(\phi_{12})^{-1}$ as follows:
{\small
\beqn    \label{jacobi-00}
 Jacobi &=& \left[ \{f,\{g,h\}\}+\{g,\{h,f\}\}+\{h,\{f,g\}\} \right] + \nonumber\\ 
        & & \left[ ( A_2 K_1 - A_1 K_2 ) +  ( B_2 J_1 - B_1 J_2 ) +
                   (C_1 I_2 - C_2 I_1) \right] z + \nonumber\\
        & & \left[ (A_1 B_2 - A_2 B_1) F  + (C_1 A_2 - C_2 A_1) E  +
                   (B_1 C_2 - B_2 C_1) D \right] z^2.
\eeqn
}
Clearly, r.h.s. of (\ref{jacobi-00}) is equal zero since all its coefficients 
are zero according to (\ref{jacobi-help1}).

\sect{Applications}


  One important class of constrained dynamical systems is characterized by 
$K$ holonomic constraints $\phi_i(q)=0$, where $i=1,\cdots,K$.  These constraints 
represent a subclass of time-independent constraints $\phi_i(q,p)=0$ considered 
in this article.  In the Dirac approach, these dynamical systems are described by 
a system of $2K$ constraints $\phi_i(q)=0$ and
$\tilde{\phi_i}(q,p)=\{\phi_i,\calH\}=0$.

  For holonomic constraints, it is convenient to introduce two $K \times K$ matrices: 
symmetric $S= (S_{ij})$ with $S_{ij}=\{\phi_i,\tilde{\phi_j}\}$ and skew-symmetric 
$A =(A_{ij})$ with $A_{ij}=\{\tilde{\phi_i},\tilde{\phi_j}\}$. The matrix $W$ 
and its inverse $C$ can then be written as 
\beqn
W=\left[\begin{array}{cc} 
              0 & S \\ -S^T & A 
        \end{array}
    \right] \,, \mbox{ and } 
C=W^{-1}=\left[ \begin{array}{cc} S^{-1} A S^{-1} & -S^{-1} \\ S^{-1} & 0 
                \end{array} \right]. 
\eeqn
  In order to compute $C$ one has to invert one symmetric $K\times K$ matrix
and do matrix multiplications twice. Symbolic computation is costly, but numerical
computation requires only $\sim K^3$ flops (floating-point operations).

  Consider now a constrained model with damping force proportional to the
generalized velocity. Such a case is described by a metriplectic structure:
\beqnn
    \{x_i, x_j \} = 0 = \{ p_i, p_j \}, ~ \{x_i, p_j\} = \delta_{ij}, \\
    <x_i, x_j > = 0, ~
    <p_i , p_j> = \delta_{ij} \lambda_i(q,p), \mbox{ where } \lambda_i \geq 0.
\eeqnn
The dissipative constraint matrix $W^{D}=\left( W_{ij}^D \right)$, where
\beqnn
  W_{ij}^D = <\tilde{\phi_i},\tilde{\phi_j}> =
           \sum_{l}\frac{\partial\tilde{\phi_i}}{\partial p_l} ~
                   \frac{\partial\tilde{\phi_j}}{\partial p_l} ~\lambda_l \,,
\eeqnn
is a symmetric $K \times K$ matrix, and let denote its inverse matrix by 
$C^{D} = (W^D)^{-1}$.
The metriplectic Dirac equations for the dynamics governed by
$\dot{f} = \{f,\calH\}_D - <f, \calH >_D$, take the form:
{\small
\beqn   \label{holo01}
   \dot{q_i} &=& \frac{\pa \calH}{\pa p_i} - \sum_{j,k=1}^{K} (S^{-1})_{j k} 
                 \frac{\pa \tilde{\phi_j}}{\pa p_i} \tilde{\phi}_k, \nonumber\\
   \dot{p_i} &=& -\frac{\pa \calH}{\pa q_i} - \sum_{j,k=1}^{K} (S^{-1})_{j k} 
                  \frac{\pa \phi_j }{\pa q_i} \{\tilde{\phi}_k,\calH\} + 
        \left[(S^{-1} A S^{-1})_{j k} \frac{\pa \phi_j}{\pa q_i} +
              (S^{-1})_{j k} \frac{\pa \tilde{\phi}_j }{\pa q_i}
        \right]  \tilde{\phi}_k             \nopagebreak  \nonumber \\
   & & - \lambda_{i} \left[ \frac{\pa \calH}{\pa p_i} - 
           \sum_{j,k=1}^K  \frac{\pa \tilde{\phi}_j }{\pa p_i} C^D_{jk} 
           \sum_{l=1}^n \lambda_l \frac{\pa \tilde{\phi}_k }{\pa p_l} \frac{\pa \calH}{\pa p_l} 
       \right].
\eeqn
}
Recursive symbolic evaluation of explicit equations for a system having $2K$ 
constraints is realized by $K$ steps.  In each step we deal with only two constraints,
e.g $\phi_i$ and $\tilde{\phi}_i$ in the $i$-th step.  In order to calculate
$2n$ explicit equations of motion subject to $2K$ constraints, i.e.
$\{x_i, \calH\}^{(K)}$ and $\{p_i, \calH\}^{(K)}$, we have to compute $(6n+3)$ 
brackets determined in $(K-1)$-th step:
$\{ x_i, \calH \}^{(K-1)}$, $\{ p_i, \calH \}^{(K-1)}$,  
$\{ x_i, \phi_{K}\}^{(K-1)}$,  $\{ x_i, \tilde{\phi}_{K} \}^{(K-1)}$, 
$\{ p_i, \phi_{K}\}^{(K-1)}$, $\{ p_i, \tilde{\phi}_{K} \}^{(K-1)}$, 
$\{ \phi_{K}, \calH \}^{(K-1)}$, $\{\tilde{\phi}_{K}, \calH \}^{(K-1)}$  
and $\{ \phi_{K}, \tilde{\phi}_{K} \}^{(K-1) }$.  

  We illustrate our procedure on the model of chain molecule often studied
in polymer and proteins physics, paying particular attention to the implementation
of the code for Dirac brackets in symbolic computer algebra system.

  A chain molecules is a constrained system consisting of  $N$ massive points
(or spherical balls) attached by rigid massless bonds having fixed length,
in $d$-dim space. We are interested in the cases when $d=2$ (planar) or $3$.
  The molecules interact with each other through a pair potential which depends
only on the distance between molecules, e.g the Coulomb interaction and/or
Lennard-Jonnes potential 
$V_{ij}= \nopagebreak 
 a \frac{q_i q_j}{r_{ij}} + 
   \varepsilon \left[ \left(\frac{\sigma_{ij}}{r_{ij}} \right)^6
                     -\left(\frac{\sigma_{ij}}{r_{ij}} \right)^{12}\right]$,
and with an external field $\tilde{U}(\vec{r}_i)$.
In a real application such a chain is immersed into a fluid matrix, thus each 
of its molecules is subject to an additional frictional force.

  We denote the position of the $i$-th molecule as $\vec{r}_i$ and its 
momentum as $\vec{p}_i$. We will lump all the positions into one vector 
$\vec{r}=(\vec{r}_1,\cdots,\vec{r}_N)$ and similarly 
$\vec{p}=(\vec{p}_1,\cdots,\vec{p}_N)$. 
It is convenient also to use the following notation: the relative position of 
$i$-th and $j$-th molecule $\vec{r}_{ij} = \vec{r}_i - \vec{r}_j$, 
the relative position of two consecutive molecules (or shortly link vector) 
$\triangle \vec{r}_{i}= \vec{r}_i - \vec{r}_{i+1}$,
the relative velocity of two consecutive molecules
$\triangle \vec{v}_{i}= \frac{\vec{p}_i}{m_i} - \frac{\vec{p}_{i+1}}{m_{i+1}}$,
and the unit vector of the link vector 
$\vec{e}_i = \frac{\triangle \vec{r}_i}{|\triangle \vec{r}_i|}$. 
\begin{figure}[htb]
\begin{center}
\centerline{\epsfxsize 0.25 \linewidth \epsfbox{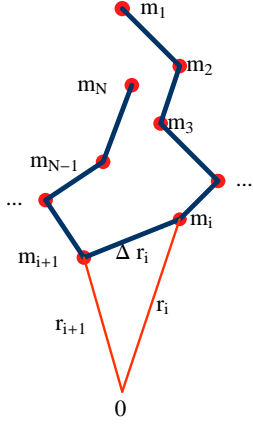} }
\caption{Linear polymer consists of $N$ molecules interacting each other.} 
\label{polymer}
\end{center}
\end{figure}
The Hamiltonian for our model reads then
\beqn 
   \calH(\vec{r},\vec{p}) = \sum_{i=1}^N \left[ \frac{|\vec{p}_i|^2}{2 m_i} +  
                                                \tilde{U} (\vec{r}_i) \right] 
                            + \sum_{j>i+1}^{N}  V_{ij}( r_{ij} ) 
                       = \sum_{i=1}^N \frac{|\vec{p}_i|^2}{2 m_i} + U(\vec{r}) \,.  
\eeqn
Putting $K=(N-1)$, the $2K$ constraints follow:
\beqn \label{constraints-polymer0}
    \phi_k(\vec{r})  = \frac{1}{2} (|\triangle \vec{r}_{k}|^2 - l_{k}^2) = 0, \; \;
    \tilde{\phi_k}(\vec{r},\vec{p}) = \triangle \vec{v}_k \cdot \triangle \vec{r}_k = 0 \,. 
\eeqn

Using this notation we can easily evaluate matrix coefficients for all the matrices 
in Eq. (\ref{holo01}). We found it convenient to collect them in the table 
\ref{tablica1}, where the $b_i$, $c_i$, $a_i$, $b_i^{(D)}$ and $c_i^{(D)}$  
(for isotropic friction $\lambda_{i(d-1) + 1} = \cdots =\lambda_{i(d-1)+d}= \Lambda_i$ 
which is the frictional coefficient for $i$-th molecule) are given as
{\small
\beqnn
    b_i = - \frac{\triangle \vec{r}_i \cdot \triangle \vec{r}_{i+1}}{m_{i+1}}
        = \frac{l_i l_{i+1}}{m_{i+1}} \cos(\alpha_i) \,, \mbox{ where } \;
          \cos(\alpha_i) = - \vec{e}_i \cdot \vec{e}_{i+1}  \,,\nonumber\\
    c_i = \frac{(m_i+m_{i+1})}{m_i m_{i+1}} |\triangle \vec{r}_i|^2
        = \left( \frac{1}{m_i}+\frac{1}{m_{i+1}} \right) l_i^2     \,, ~~ 
    a_i = \frac{ \triangle \vec{r}_i \cdot \triangle \vec{v}_{i+1} 
              -\triangle \vec{v}_i \cdot \triangle \vec{r}_{i+1} }{m_{i+1}} \,, \\
    c^{(D)}_i = \left( \frac{\Lambda_i}{m_i^2} + 
                       \frac{\Lambda_{i+1}}{m_{i+1}^2} \right) l_i^2 \,, ~~
    b^{(D)}_i = \frac{- \Lambda_{i+1} }{m_{i+1}^2} \triangle \vec{r}_i \cdot 
                                                  \triangle \vec{r}_{i+1} = 
                \frac{ \Lambda_{i+1} }{m_{i+1}^2} l_i l_{i+1} \cos(\alpha_i) \,. \nonumber 
\eeqnn
}
\begin{table}[ht] 
\beqnn
\begin{array}{|c|c|c|c|}  
      \hline 
                     &    &    &   \\   
    \mbox{Condition} &  S_{ij}=\{\phi_i,\tilde{\phi}_j\}  
                             &  A_{ij}=\{\tilde{\phi}_i,\tilde{\phi}_j\} 
                             &  S^{(D)}_{ij}= < \tilde{\phi}_i,\tilde{\phi}_j > \\ 
      \hline
          |i-j| > 1  &  0   &  0  &  0  \\  
      \hline
          j = i+1    & b_i  & a_i &  b_i^{(D)} \\ 
      \hline
          j = i      & c_i  &  0  &  c_i^{(D)}  \\ 
      \hline
          j = i - 1  & b_j  &  -a_j  &  b^{(D)}_j \\ 
      \hline
\end{array}
\eeqnn
\caption{Elements of the matrices $S$, $A$ and $S^{(D)}$}   
\label{tablica1}
\end{table} 
Thus, the matrices $S$, $S^{(D)}$ are symmetric tridiagonal, while $A$ is 
skew-symmetric tridiagonal, shown in the table \ref{sym-anti-mat}.
\begin{table}[ht]
{\small
\beqnn
S= \left[  
     \begin{array}{ccccc}
       \displaystyle   c_1 & b_1 & 0 & \cdots & 0 \\ 
       \displaystyle   b_1 & c_2 & b_2 & \ddots & \vdots \\
       \displaystyle     0 & \ddots & \ddots & \ddots & 0 \\
       \displaystyle \vdots & \ddots & \ddots & \ddots & b_{K-1}\\  
       \displaystyle    0  & \cdots &   0    & b_{K-1} & c_K
     \end{array}
   \right] \mbox{ and } 
A= \left[  
      \begin{array}{ccccc}
        \displaystyle  0 &  a_1 &  0  & \cdots & 0  \\ 
        \displaystyle  -a_1 & 0 & a_2 & \ddots & \vdots \\
        \displaystyle  0 & \ddots & \ddots & \ddots & 0  \\
        \displaystyle  \vdots & \ddots  & \ddots & \ddots & a_{K-1}\\  
        \displaystyle  0   & \cdots &  0  & -a_{K-1} & 0
      \end{array}
   \right] 
\eeqnn
}
\caption{\label{sym-anti-mat}Symmetric and skew-symmetric Tridiagonal Matrices 
         $S$ and $A$}
\end{table}
   For homogeneous polymer in homogeneous environment, consisting of identical 
molecules, $l_i=l$ and $m_i = m$, all formulas on elements of $S, S^{(D)}$ 
become even simpler:
{\small
\beqn
   c_i = \frac{2 l^2}{m}, \; 
   b_i = \frac{l^2}{m} \cos(\alpha_i), \mbox{ and } \;
   c^{(D)}_i = \frac{2 \Lambda l^2}{m^2}, \; 
   b^{(D)}_i = \frac{\Lambda l^2}{m^2} \cos(\alpha_i). 
\eeqn
}
  Though the tridiagonal matrices have been considered numerically for years, 
the explicit analytic formulas for elements of the inverse matrix of a 
tridiagonal matrix are known only in some special cases \cite{Hu-Connell}: 
$b_i=b$ and $c_j=c$. 
Here we propose a general expression for elements of $S^{-1}$. Details of the 
derivation of that formula are given in the Appendix A.

  Let $S(1, \cdots, i-1)$ be the top left $(i-1) \times (i-1)$ matrix containing 
rows and columns $\{1,\ldots,i-1 \}$ of $S$ and $S(j+1,\cdots,K)$ be the bottom 
right $(K-j) \times (K-j)$ matrix containing rows and columns $\{ j+1,\ldots,K \}$ 
of $S$, we get the following recursive formula:  
\beqn  \label{imp-invS}
   (S^{-1})_{i,j} = (-1)^{i+j} \frac{ |S(1, \cdots, i-1)| |S(j+1, \cdots, K)| }
                                 { |S(1 \cdots K)| }   
   b_i b_{i+1} \cdots b_{j-1} \,,  
\eeqn
for $i \leq j$, and $S^{-1}$ is symmetric.  Since both matrices $S$ and $S^{(D)}$ 
have a similar form, we can use the formula (\ref{imp-invS}) in calculating their 
inverse.

  Furthermore, for $K \geq n > l \geq 1$, the $|S(l,\cdots,n)|$ is calculated 
from the recursive relation: $|S(\emptyset)|=1,~ |S(l)|=c_l$, 
$|S(l, \cdots, n)|=c_n |S(l, \cdots, n-1)| - b_{n-1}^2 |S(l, \cdots, n-2)|$.

  With the formula (\ref{imp-invS}), it is easy to show that the inverse matrix 
of a symmetric tridiagonal matrix is one-pair matrix.
Numerically it can be computed fast with $O(N)$ complexity cost, and with modest 
memory usage.
  Since the recursion relation (\ref{imp-invS}) is rather involved, we can only 
calculate the Dirac equations via recursion. More technical details are presented
in our paper posted on the arxiv page.

\underline{{\bf Discussion}}

  We have implemented our formalism using the package Mathematica version 
5.2 and 6.0, the computer algebra system, both for symbolic and numerical 
calculation, and measured the CPU time needed in computing explicit analytical 
r.h.s. of (\ref{holo01}) in two ways: one based on the formula (\ref{imp-invS}) 
and the other based on the recursion relation (\ref{recursive-anti00}).
All computation have been done on an ordinary PC (with dual core processor
$1.6$ GHz and 1GB RAM) running MS Windows XP and Linux FC6.
\begin{figure}[htb]
\begin{center}
\centerline{\epsfxsize 0.8\linewidth \epsfbox{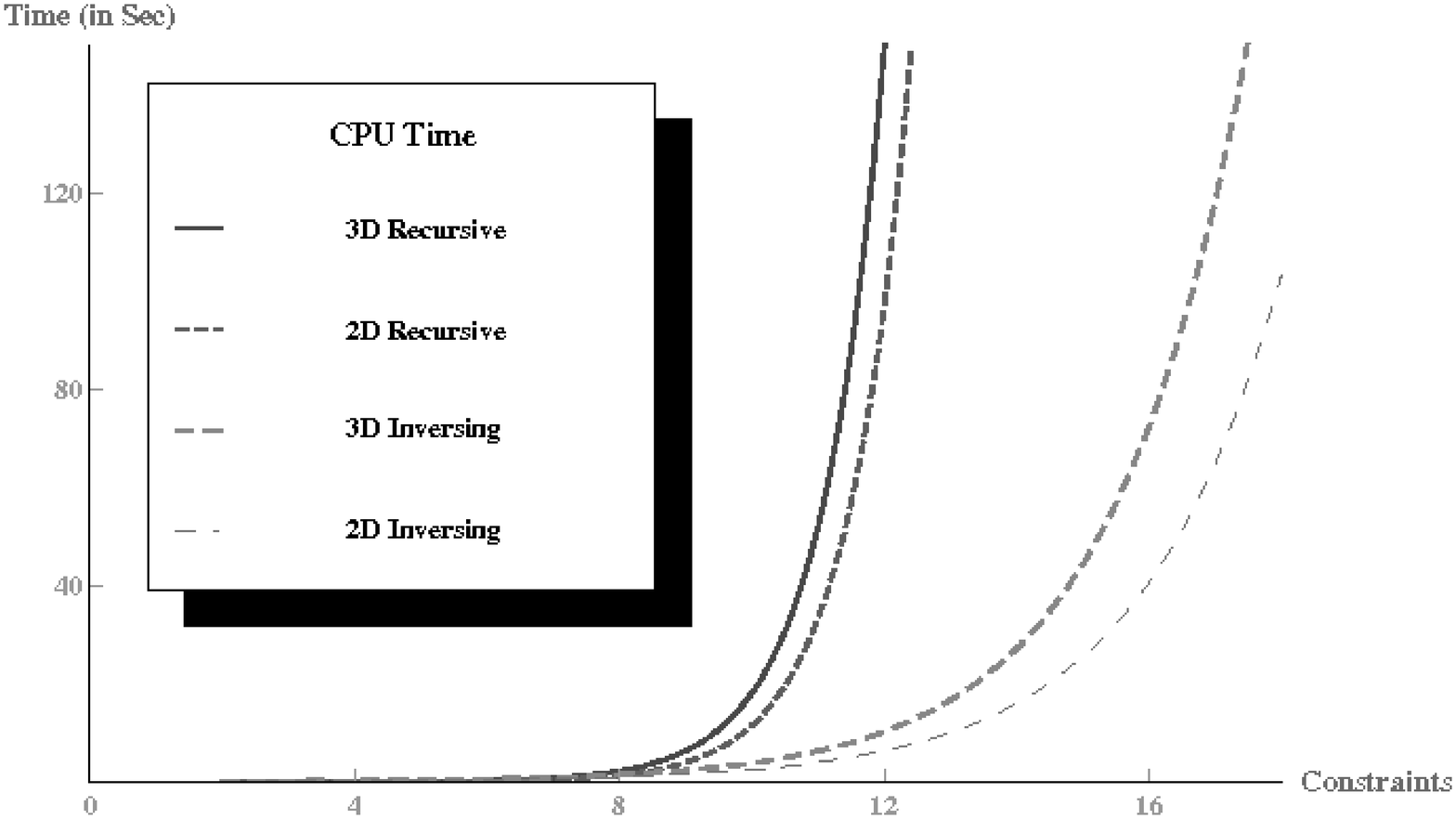}}
\end{center}
\end{figure}
  The symbolic computing time for one pair of equations in $3$-dim, after using 
least square interpolation, seems to grow with the number of constraints like
$~0.028 \, e^{0.49 K}$ and as $~0.00046 \, e^{1.06 K}$ for method inverting
triangular matrices and using recursive formula, respectively.
  Consequently, the recursive formula is reasonably good only for systems with 
less than $12$ constraints.  Since the computing time in both methods grow 
exponentially in the number of constraints, computing explicit analytical 
Dirac equations seems to be inapplicable for very long chains.
  However, fast algorithm for numerical inversion of tridiagonal matrices does 
exist and has a complexity $O(N)$.  Thus, Dirac finite difference equations for 
long chains are computable.

  Having explicit equations of motion one can solve them numerically 
either by standard explicit/implicit Runger-Kutta algorithm or standard 
Mathematica's ODE solver {\em NDSolve}.

   Another important issue is that alternatively to the system of equations 
(\ref{holo01}), one can consider the following system:
{\small
\beqn   \label{holo01-simple}   
   \dot{q_i} &=& \frac{\pa \calH}{\pa p_i}, \nopagebreak\\ 
   \dot{p_i} &=& -\frac{\pa \calH}{\pa q_i} - 
   \sum_{j,k=1}^{K} (S^{-1})_{j k} \frac{\pa \phi_j }{\pa q_i} 
        \{\tilde{\phi}_k ,\calH\} -
   \lambda_{i} \left[ \frac{\pa \calH}{\pa p_i} - 
         \sum_{j,k=1}^K  \frac{\pa \tilde{\phi}_j }{\pa p_i} C^D_{jk} 
         \sum_{l} \lambda_l \frac{\pa \tilde{\phi}_k }{\pa p_l} \frac{\pa \calH}{\pa p_l} 
     \right] \,. \nonumber
\eeqn
}

   Since constraints are Casimir elements regarding to Dirac bracket, any
solution of (\ref{holo01}) with initial conditions satisfying all constraints,
automatically 
satisfies all constraints for all time. Therefore it must also be a solution of (\ref{holo01-simple}).

\begin{figure}[htb]
\begin{center}
\subfigure[Energy calculated from eq. (\ref{holo01}) and (\ref{holo01-simple})] 
{\includegraphics[scale=0.25]{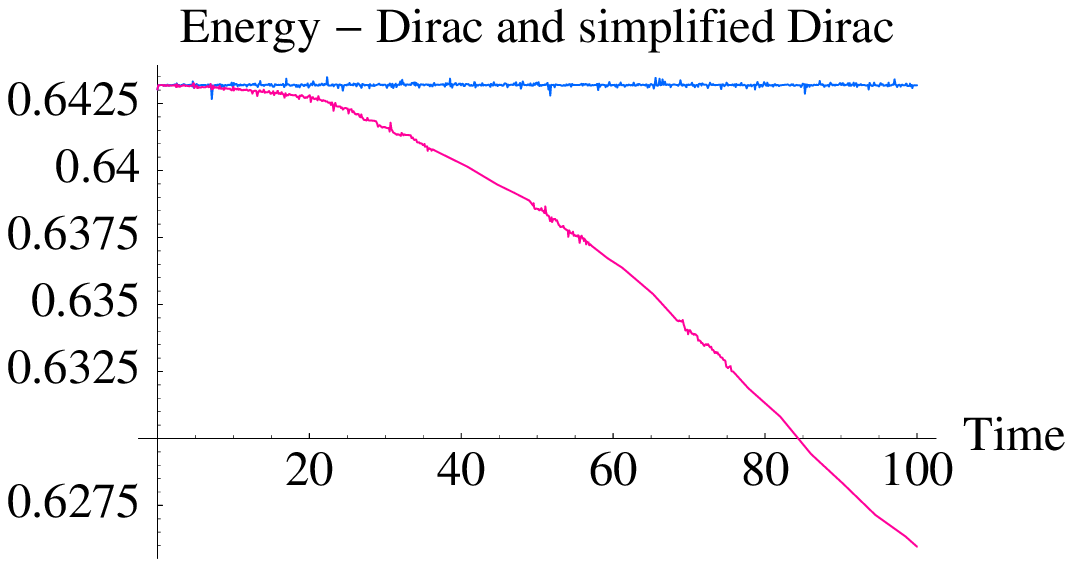}}
\subfigure[Energy calculated from (\ref{Lagrange_multiplier_eqs})]
{\includegraphics[scale=0.23]{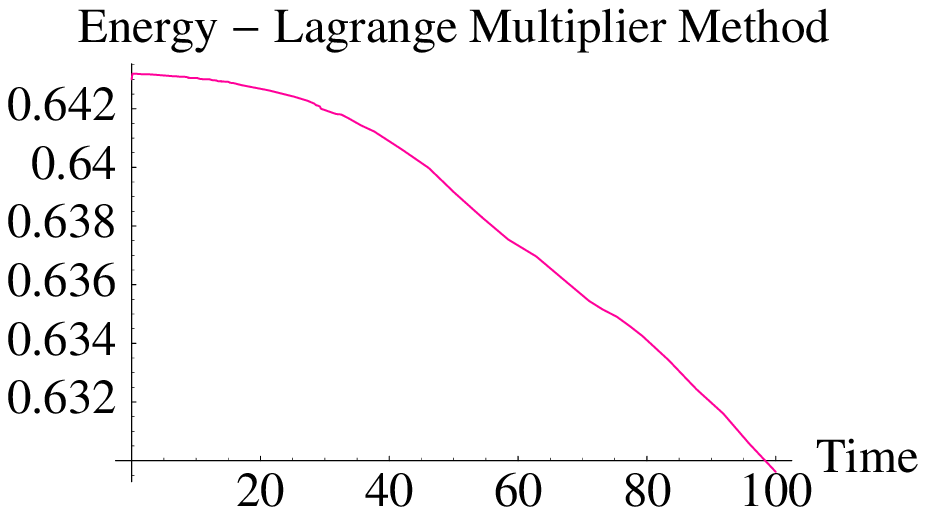}}
\subfigure[Sum of constraints errors calculated from eq. (\ref{holo01})]
{\includegraphics[scale=0.236]{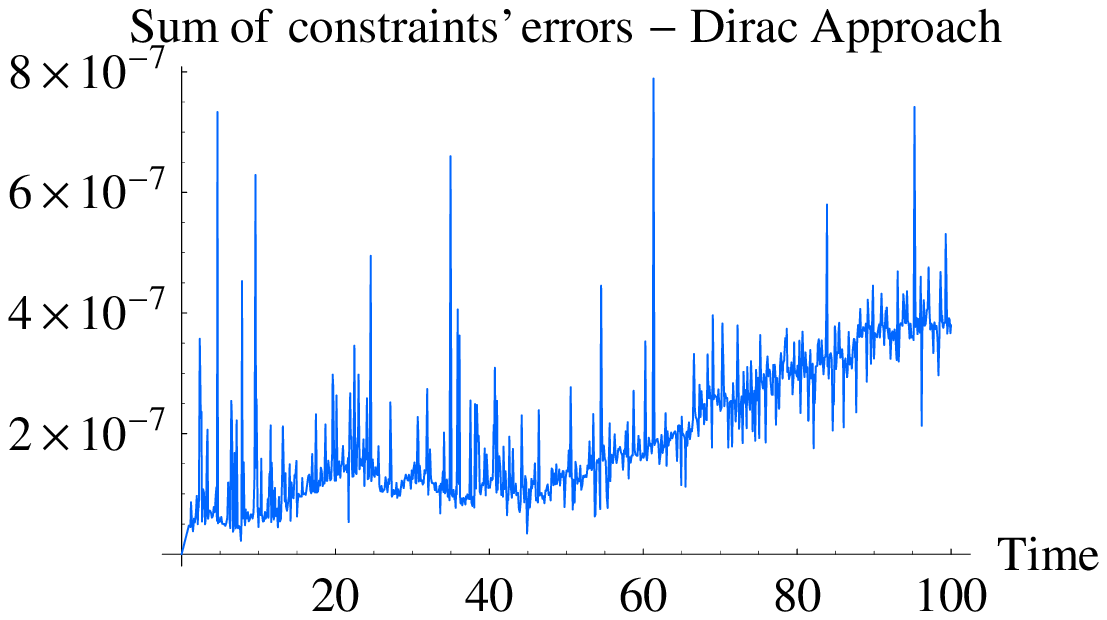}}
\subfigure[Sum of constraints errors calculated from eq. (\ref{Lagrange_multiplier_eqs})]
{\includegraphics[scale=0.236]{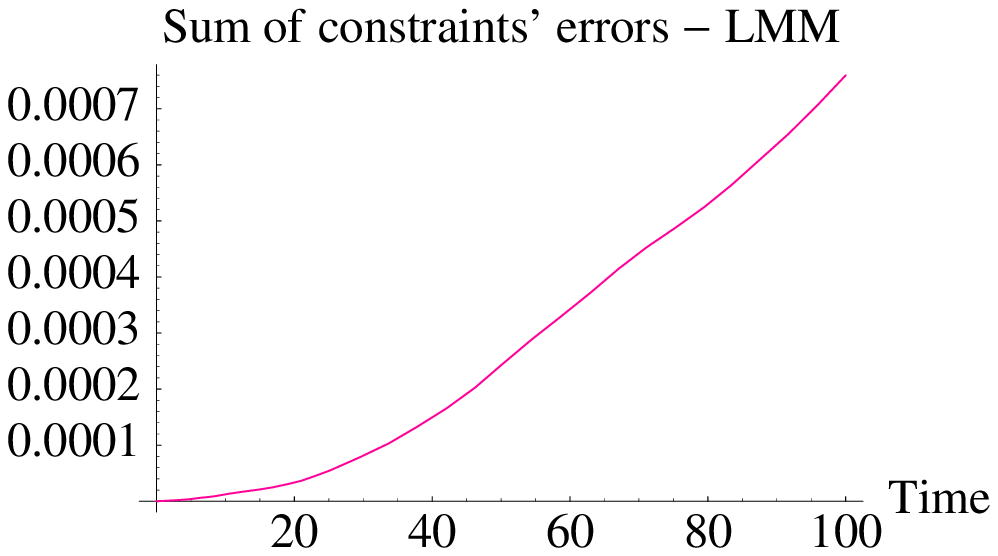}}
\end{center}
\caption{Numerical test: Energy and Constraints errors for
          4-pendulums dynamics described by the Hamilton-Dirac eq. (\ref{holo01}), 
          simplified Dirac (\ref{holo01-simple}) and LMM (\ref{Lagrange_multiplier_eqs}) 
          using default numerical algorithm NDSolve.
         For simplicity we have chosen a system consisting of $4$ equal masses 
         which are in the axis $x$ at the beginning, and whose initial velocities 
         have random values satisfying constraints' equations.}
\label{energy-conservation}
\end{figure}

   This fact and the uniqueness of solution (locally) imply that two systems
(\ref{holo01}) and (\ref{holo01-simple}) are equivalent.
In our tests, symbolic computation for the latter is 6-7 times faster than for
the former.  Moreover, for non-dissipative mechanical systems, the latter is 
exactly the system of equations obtained from the {\it Lagrange Multiplier Method} 
(LMM), eq. (\ref{Lagrange_multiplier_eqs}) in the Appendix \ref{app-B}.
   Though these two systems are mathematically equivalent, they are not equivalent
for numerical algorithms approximating solution, which means that errors grow
differently for each of them even if using a common numerical algorithm.  Errors in
computing approximate solution of the {\it LMM-like} eq. (\ref{holo01-simple}) or (\ref{Lagrange_multiplier_eqs}), always grow faster than those of the {\it Dirac-like} 
eq. (\ref{holo01}).
   We studied numerically the violation of energy and bond length constraints
for a particular polymer with one fixed end, eg. $N$-pendulum described in
the Appendix \ref{appendix-N-pendulums}.  These numerical results are presented
briefly in the figure \ref{energy-conservation}.
  In summation, standard numerical algorithms seem to work well with Dirac-like 
equations.  To deal numerically with LMM-like equations, we recommend to use either
constrained algorithms (eg. SHAKE, LINCS) or other advanced symplectic/poisson ones,
which have been developed recently.


  Although in the simulation, polymers with nearly constant bond length, 
called stiff bead-spring chains, are more often considered than those 
with rigid constant length, named bead-rod chains, the matrix $S$ which 
has been carefully studied here, is closely related to the metric potential 
$U = \frac{1}{2} kT \log(| S |)$ in the statistical mechanics of Polymers 
\cite{Fixman}.

  The application of bracket formalism to the non-linear many particle models 
is possible by time consuming.  We have looked at the possibility of using our 
method to obtain a set of analytical equations and simulate mechanics of the 
caricatured human body \cite{Badler-99}.

\begin{figure}[htb]
\begin{center}
\centerline{\epsfxsize 0.2 \linewidth \epsfbox{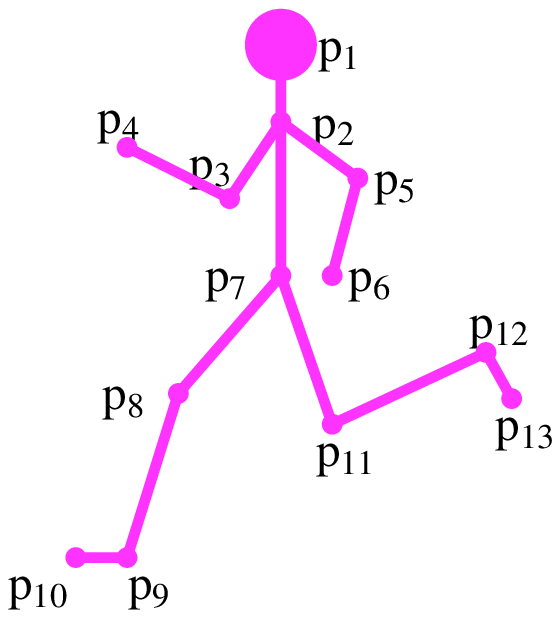} 
            \epsfxsize 0.16 \linewidth \epsfbox{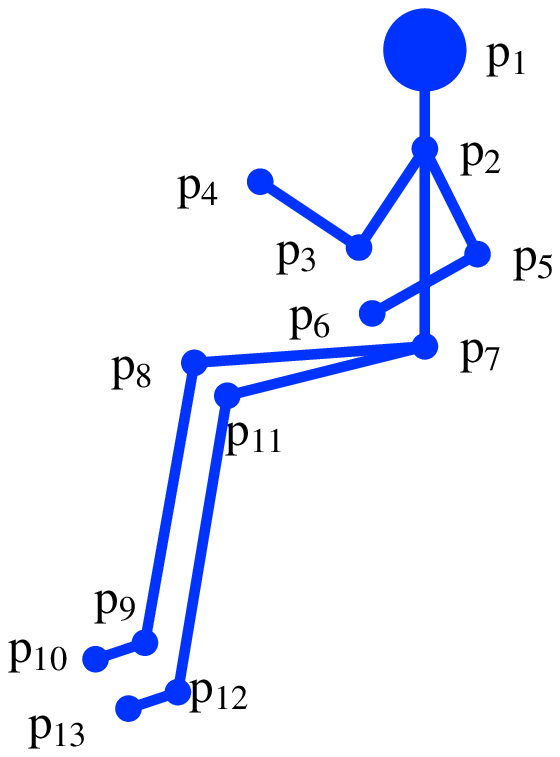}} 
\caption{Humanoid is a (dissipative) constrained dynamical system with 
         $24$ phase space constraints. This is an example of non-linear chain.} 
\label{humanoid}
\end{center}
\end{figure}

  Instead of models for body dynamics such as inverted pendulum \cite{Maurer-05}, 
or elastic string \cite{Leuk-98} are used, we used skeletal humanoid consisting 
of 13 material points, fig. \ref{humanoid}.
  We found that symbolic calculation each pair of explicit analytical equations for 
humanoid takes app. 9 minutes using formula (\ref{imp-invS}) for inverting matrix 
$S$, of uninterrupted Mathematica performance in PC.

\sect{Conclusions}

In this article we have reviewed a geometric construction of Dirac-like 
brackets and proved recursive character of such brackets. We showed that 
computing explicit dynamical equations based on these brackets may be 
difficult, but it is possible to produce analytical equations even for 
systems with many constraints.

We have applied here the Dirac procedure for metriplectic mechanical models 
with finite degrees of freedom, but in our previous work we have shown its 
usefulness for continuous models \cite{SNT3}, for example incompressible 
hydrodynamics \cite{SNT1}.
Fixman \cite{Fixman} have used constraints approach in formulation of 
statistical mechanics of various polymer models.  The fact that constraints 
can then be visualized as a kind of temperature dependent potential is not 
unusual.
Fixman and others have restricted their procedure to the equilibrium calculations.
Our formalism allows us to go beyond the equilibrium application and see the form
of the constrained Liouville equations, modifications in the dynamical modes 
coupling due to the presence of constraints and possible the role of the constraints 
play in removing the singularities appearing in low dimensional systems statistical 
mechanics.
For example, the fact that the transport coefficients, like viscosity, thermal 
conductivity and diffusion coefficient do not exists in $d=2$, can be modified 
by presence of the constraints in a fashion analogous to that mentioned in 
\cite{Turski2d}.
\ack
The work one of us (SN) was partially supported by the Hanoi University of Science
Grant No. TN-08-15 and the other (LAT) was partially supported by the Polish Ministry
of Science and Higher Education Grant No. N20204232/1171.
\appendix
\section{Lagrange Multiplier Method} \label{app-B}
The purpose of this section is to show that computing explicit analytical equations 
in the Lagrangian formalism is equally difficult as in the Dirac formalism.

For simplicity, suppose that all constraints of the form: $\phi_k(q)=0$, 
$k=1,\ldots, K$ and $q=(q_1,\ldots,q_n)$.
Lagrangian of constrained system is a sum of unconstrained Lagrangian and a linear
combination of constraints: 
$\calL(q,\dot{q}) = \calL_0(q,\dot{q}) - \sum_{k=1}^K \lambda_k \phi_k(q)$.
The Euler-Lagrange equations read 
\beqnn
  \frac{\pa}{\pa t} \left(\frac{\pa \calL}{\pa \dot{q}} \right) - \frac{\pa \calL}{\pa q} = 0 \,.
\eeqnn
Suppose Lagrangian of the form 
$\calL_0 = T(\dot{q}) - V(q) = \frac{1}{2} \, \dot{q}^T \, M \, \dot{q} - V(q)$, 
with introducing conservative force $F = - \frac{\pa V}{\pa q}$, 
the Euler-Lagrange equations become
\beqn \label{lag1}
    M \ddot{q}  &=& F - \sum_{k=1}^K \lambda_k \frac{\pa \phi_k}{\pa q_i}  = 
                    F - B \lambda  \,,  
\eeqn
where $B = (B_{ik})$ is a $n \times K$ matrix whose elements 
$B_{ik} = \frac{\pa \phi_k}{\pa q_i}$. 
Since $\phi_k(q)=0$, all first and second time derivatives of $\phi_k$ vanish: 
\beqn
    0 = \frac{d \phi_k}{d t} = \sum_{i=1}^n \frac{\pa \phi_k}{\pa q_i} \dot{q}_i  
    ~ \mbox { or }~  [ B^T \dot{q} ]_k = 0 \,,  \\
    \label{lag2} 
    0 = \frac{d^2 \phi_k}{d^2 t} 
      = \sum_{i,j=1}^n \frac{\pa^2 \phi_k}{\pa q_i \pa q_j} \dot{q}_i \dot{q}_j +
        \sum_{i=1}^n   \frac{\pa \phi_k}{\pa q_i} \ddot{q}_i
      = G_k + [B^T \ddot{q}]_k  \,,   
\eeqn
where $G_k = \sum_{i,j=1}^n \frac{\pa^2 \phi_k}{\pa q_i \pa q_j} \dot{q}_i \dot{q}_j$.
Substituting for $\ddot{q} = M^{-1} [ F - B \lambda ]$, derived from (\ref{lag1}), 
in (\ref{lag2}) we get:  
\beqn
    0 &=& G + B^T M^{-1} [ F - B \lambda] \,,     
\eeqn 
here $G=(G_k), ~ \lambda=(\lambda_k)$ are column vectors $K \times 1$ and 
$F=(F_j)$ is a column vector $n \times 1$.  Therefore, 
$[G + B^T M^{-1} F] = (B^T M^{-1} B) \lambda$ or 
$\lambda = (B^T M^{-1} B)^{-1} [G + B^T M^{-1} F]$.  
Substituting this back to (\ref{lag1}) we get explicit constrained equations:
\beqn \label{Lagrange_multiplier_eqs}
     M \ddot{q}  &=& F - B (B^T M^{-1} B)^{-1} [G + B^T M^{-1} F] \,.   
\eeqn
Thus, for achieving explicit equations in the Lagrangian formalism, it is also 
necessary to compute analytical inversion of the $K \times K$ matrix $(B^T M^{-1} B)$ 
which is exactly equal the matrix $S$ in the Dirac approach where the Hamiltonian 
obtained from the Legendre transformation: $\calH =  p \dot{q} - \calL$ with 
$p = \frac{\pa \calL}{\pa \dot{q}}$.
\section{N-pendulum in d dimensional space} \label{appendix-N-pendulums}
We denote the position of the $i$-th mass as 
$\vec{r}_i=(x_{d(i-1)+1},\ldots,x_{d i})$, 
its momentum as $\vec{p}_i=(p_{d(i-1)+1},\ldots,p_{d i})$,
the relative position of $i$-th and $j$-th mass 
$\vec{r}_{ij} = \vec{r}_i - \vec{r}_j$, 
the relative position of two consecutive masses (or shortly link vector) 
$\triangle \vec{r}_{i}= \vec{r}_i - \vec{r}_{i+1}$,
the relative velocity of two consecutive masses
$\triangle \vec{v}_{i}= \frac{\vec{p}_i}{m_i} - \frac{\vec{p}_{i+1}}{m_{i+1}}$,
and the unit vector of the link vector 
$\vec{e}_i = \frac{\triangle \vec{r}_i}{|\triangle \vec{r}_i|}$. 
\begin{figure}[htb]
\begin{center}
\centerline{\epsfxsize 0.3 \linewidth \epsfbox{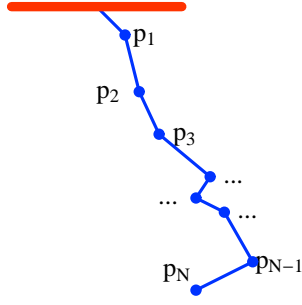}  } 
\caption{N-pendulum is a constrained system with $N$ length constraints, which 
         can be viewed as a linear polymer with fixed end.} 
\label{pendulum-pict}
\end{center}
\end{figure}
\subsection{Hamilton-Dirac description for $N$-pendulum}
The Hamiltonian is given by
$ \calH(\vec{r},\vec{p}) = \sum_{i=1}^N \left[ 
                  \frac{|\vec{p}_i|^2}{2 m_i} +  g \, m_i \, x_{d i} \right]$,
and $2N$ second-class constraints follow:
\beqn
  \phi_k(\vec{r}) &=& \left\{ 
              \begin{array}{ll} 
                \frac{1}{2} (\sum_{j=1}^d x_j^2 - {l_1}^2) 
               = \frac{1}{2} \left( | \vec{r}_1 |^2 - l_1^2 \right)  & \mbox{ for } k=1 \,,\\
                 \frac{1}{2} \left( |\triangle \vec{r}_k|^2 - {l_k}^2 \right) &   
                                \mbox{ for } 1 < k \leq N 
              \end{array} \right.      \\
    \tilde{\phi}_k(\vec{r},\vec{p}) &=& \{\phi_k,\calH \} 
          = \left\{ 
                \begin{array}{ll} 
                    \vec{r}_1 \cdot \vec{v}_1  & \mbox{ for } k=1 \,,\\
                    \triangle \vec{v}_k  \cdot \triangle \vec{r}_k & \mbox{ for } 1 < k \leq N \,.
                \end{array} \right.
\eeqn
\subsection{Lagrange Multiplier Method for $N$-pendulum}
The Lagrangian is given by
$    \calL(\vec{r},\vec{p}) =  
    \sum_{i=1}^N \left[ \frac{|\vec{p}_i|^2}{2 m_i} -  g \, m_i \, x_{d i} \right]$,
and $N$ length-constraints follow:
\beqn
    \phi_k(\vec{r}) &=& \left\{ 
       \begin{array}{ll} 
            \frac{1}{2} [\sum_{j=1}^d x_j^2 - l_1^2]                 
            = \frac{1}{2} \left( | \vec{r}_1 |^2 - l_1^2 \right)  & \mbox{ for } k=1 \,,\\
            \frac{1}{2} (|\triangle \vec{r}_k|^2 - {l_k}^2) & \mbox{ for } 1 < k \leq N \,.
       \end{array} \right.
\eeqn
In order to calculate explicit eq. (\ref{Lagrange_multiplier_eqs}) we need to 
calculate explicit elements of $S^{-1}$ where $S$ follows: 
{\small
\beqn \label{appendix-S}
B^T M^{-1} B = S = 
  \left[  
  	\begin{array}{cccccc}
  	        \displaystyle    c_1 & b_1 & 0 & \cdots & \cdots & 0 \\ 
  	    	\displaystyle    b_1 & c_2 & b_2 &  0  &        & \vdots \\
 	        \displaystyle     0  & b_2 & c_3 & b_3 & \ddots & \vdots \\
      		\displaystyle \vdots & \ddots & \ddots & \ddots & \ddots & 0 \\
      		\displaystyle \vdots &        & \ddots & \ddots & c_{N-1} & b_{N-1}\\  
      		\displaystyle     0  & \cdots & \cdots &   0    & b_{N-1} & c_N
     	\end{array}
  \right] \,,
\eeqn
}
here 
\beqn
    b_i = \left\{ 
          \begin{array}{ll} 
                \frac{ \vec{r}_1 \cdot \triangle \vec{r}_1}{m_1}  
                = \frac{l_1 l_2}{m_1} \cos(\alpha_1)  & \mbox{ for } i =1 \,,\\  
         	- \frac{\triangle \vec{r}_{i-1} \cdot \triangle \vec{r}_i}{m_{i}} 
              = \frac{l_i l_{i+1}}{m_i} \cos(\alpha_i) & \mbox{ for } 1 < i \leq N-1 \,,        
           \end{array} 
           \right.  \nonumber \\
    c_i = \left\{ 
          \begin{array}{ll}
                \frac{1}{m_1} {l_1}^2 & \mbox{ for } i =1 \,,\\     
         	\left( \frac{1}{m_{i-1}}+\frac{1}{m_i} \right) l_i^2  & 
                           \mbox{ for } 1 < i \leq N \,.
          \end{array} 
          \right.   
\eeqn

\section{Symbolic Inversion of Symmetric Tridiagonal Matrices}

In this section we discuss problem of symbolic inversion general 
symmetric tridiagonal matrix whose explicit form is given in  
(\ref{sym-tri-mat}).
{\small
\beqn \label{sym-tri-mat}
S= \left[
     \begin{array}{cccccc}
      \displaystyle    c_1 & b_1 & 0 & \cdots & \cdots & 0 \\ 
      \displaystyle    b_1 & c_2 & b_2 &  0  &        & \vdots \\
      \displaystyle     0  & b_2 & c_3 & b_3 & \ddots & \vdots \\
      \displaystyle \vdots & \ddots & \ddots & \ddots & \ddots & 0 \\
      \displaystyle \vdots &        & \ddots & \ddots & c_{K-1} & b_{K-1}\\  
      \displaystyle     0  & \cdots & \cdots &   0    & b_{K-1} & c_K
     \end{array}
   \right]
\eeqn }
Notation: Let $M=(M_{i,j})$ be a matrix.  Define as $M(i_1,\ldots,i_p;j_1,\ldots,j_q)$ 
the matrix consisting of elements $M_{i,j}$ where $i \in \{i_1,\ldots,i_p\}$ and 
$j \in \{j_1,\ldots,j_q\}$.  In the case $\{i_1,\ldots,i_p\} \equiv \{j_1,\ldots,j_q\}$ 
instead writing $M(i_1,\ldots,i_p;i_1,\ldots,i_p)$ we will write $M(i_1,\ldots,i_p)$.  
\begin{defi}  \label{def-one-pair-matrix}
	A $n \times n$ symmetric matrix $Q$ is called an one-pair matrix if 
	its elements are products of components of two vectors $u=(u_1,\ldots,u_n)$ 
	and $w=(w_1,\ldots,w_n)$, i.e. 
\beqn  \label{one-pair-matrix}
     Q_{i,j} = \left\{  \begin{array}{cc}
                             u_i w_j  & \mbox { for } i \leq j \\
                             u_j w_i  & \mbox { for } i \geq j \,.
                        \end{array}   \right.     
\eeqn
\end{defi}
\subsect{Direct computation}
Since $(S^{-1})_{ij} = (-1)^{i+j} |S^{(j;i)}| / \det S $, in order to compute 
elements of $S^{-1}$ one has to compute the determinant $|S|=\det S$ and the 
co-factor $(-1)^{i+j} |S^{(j;i)}|$.  
First, if denote the determinant of $k \times k$ symmetric tridiagonal matrix by 
$S_k$, then $S_k$ can be calculated recursively as follows: 
\beqn
    S_0 = 1, \;  S_1 = c_1, \;  S_k = c_k S_{k-1} - b_{k-1}^2 S_{k-2}.
\eeqn  
Second, since $S$ is tridiagonal, the $S^{(j;i)}$ has three decoupled sub-blocks 
on the main diagonal with the order $(i-1)$, $|j-i|$ and $(K-j)$. 

Denote the determinant of a matrix $(S_{ab})$ whose indexes $a,b$ belong 
to the set $\{i_1,i_2,\ldots,i_p\}$ by $|S(i_1, i_2, \ldots, i_p)|$, 
with $i \leq j$ we have
\beqn  \label{imp-invS-appendix}
   (S^{-1})_{i,j} &=& (-1)^{i+j} \frac{ |S(1, \ldots, i-1)| |S(j+1, \cdots, K)| }
                                 {|S(1, \ldots, K)|} ~   
                       b_i b_{i+1} \ldots b_{j-1} \,.  
\eeqn
For $K \geq k \geq l \geq 1$, the $|S(l,\ldots,k)|$ is computed from 
recursive relation:  $|S(\emptyset)|=1, |S(l)|=c_l$, 
$|S(l,\ldots,k)|=c_k |S(l, \ldots, k-1)| - b_{k-1}^2 |S(l,\ldots,k-2)|$.

Introducing $d_i^{(l)}=|S(l,\ldots,l+i-1)|$ with $l+i-1 \leq K$, the sequence 
$d_i^{(l)}$ for fixed value of $l$ is determined by the recursion: 
\beqn \label{recursion-appendix01}
     d_0^{(l)}=0, \;\; 
     d_1^{(l)} = c_l, \;\;
     d_{i+1}^{(l)}=c_{l+i} d_i^{(l)} - b_{l+i-1}^2  d_{i-1}^{(l)} \,.
\eeqn
The elements of the inverse matrix are calculated as follows: 
\beqn  \label{imp-invS-appendix01}
   (S^{-1})_{i,j} &=& (-1)^{i+j} \frac{ d_{i-1}^{(1)} d_{K-j}^{(j+1)} }
                                 {d_K^{(1)}}   ~
                      b_i b_{i+1} \cdots b_{j-1} \, , \mbox{ for $i \leq j$ }.
\eeqn
If $b_i \neq 0$, one can introduce  
\beqn
  u_i &=&\frac{(-1)^i}{| S |} |S(1,\ldots, i-1)| ~ b_i b_{i+1} \cdots b_{K-1} \\     
  w_j &=& (-1)^j \frac{|S(j+1, \ldots, K)|}{b_j b_{j+1} \cdots b_{K-1}} \,,
\eeqn
and express elements of $S^{-1}$ by two vectors $(u_k)$ and $(w_k)$
\beqn
     (S^{-1})_{i,j} &=& u_i w_j ~ \mbox { for } i \leq j \,.
\eeqn
Thus, we have proved that the inverse of a symmetric tridiagonal matrix 
is one-pair matrix. The reverse statement remains true. 
Eq. (\ref{recursion-appendix01}) and (\ref{imp-invS-appendix01}) seem to define
the most effective algorithm for computing elements of $S^{-1}$. 
\subsect{Block diagonalization}
This method based on the observation that for a symmetric tridiagonal matrix 
$S$ it is easy to find a sequence of upper triangular (non-symmetric) 
tridiagonal matrices $U_k$, $k=1,\cdots,K-1$, with the main diagonal 
$\{1,\ldots,1,x_k,z_k,1,\ldots,1\}$ and its upper neighbour diagonal 
$\{0,\ldots,0,y_k,0,\ldots,0\}$, i.e.
{\small
\beqn
U_k &=& 
\left[ \begin{array}[c]{cccccc}
     		1 &    0   & \cdots & \cdots & \cdots &   0   \\
     		0 & \ddots & \ddots &        &        & \vdots \\
           \vdots & \ddots &   x_k  &   y_k  &        & \vdots \\
           \vdots &        &    0   &   z_k  & \ddots & \vdots \\
           \vdots &        &        & \ddots & \ddots &   0   \\ 
     		0 &   0    & \cdots & \cdots &   0    &   1    
	\end{array}
\right] \,,
\eeqn  
}
such that $\forall k, ~~ 1 \leq k \leq K-1$:  
{\small
\beqn
(U_1 \ldots U_k)^T  ~S~ (U_1 \ldots U_k) &=& 
\left[ \begin{array}[c]{cccccc}
     		  I_{k+1} & \vdots   &    0     & \cdots  & \cdots &  0 \\
     		  \cdots  &          & \beta_{k+1} & \ddots  &        & \vdots \\
     		      0   & \beta_{k+1} & c_{k+2}  & b_{k+2} & \ddots & \vdots \\
                  \vdots  & \ddots   & b_{k+2}  & \ddots  & \ddots &  0  \\
     		  \vdots  &          & \ddots   & \ddots  &  \ddots & b_{K-1} \\ 
     		      0   &          & \cdots   &    0    & b_{K-1} & c_{K} 
	\end{array}
\right] \,.
\eeqn }
Thus, their product $U= U_1 \cdots U_{K-1}$ is the upper triangular satisfying: 
$U^T ~ S ~ U = I$.  This implies that the inverse matrix of $S$ is a product of $U$ 
and its transposition, $S^{-1} = U U^T$.  With convention $z_0=1$, elements of $U$ 
follow: 
\beqn \label{elements-U}
    U_{i,j} &=& 
    \left\{ 
	\begin{array}{c}
       		x_j (y_i \cdots y_{j-1}) z_{i-1}  ~ \mbox{ for } i \leq j-1 \\
       		x_{j} z_{j-1} ~ \mbox{ for } i= j \\
                0   ~ \mbox{ for } i > j \,.
	\end{array}
   \right.
\eeqn  
In order to calculate $U_k$ one has to solve recursively a system of 3 quadratic 
equations
\beqn \label{quadratic-system}
\left[ \begin{array}[c]{cc}
    		x_k & 0 \\
    		y_k & z_k  
	\end{array}
\right] 
\left[ \begin{array}[c]{cc}
    		a_{k} & \beta_{k} \\
    		\beta_{k} & d_{k}  
	\end{array}
\right] 
\left[ \begin{array}[c]{cc}
    		x_k & y_k \\
    		0 & z_k  
	\end{array}
\right] 
 &=& I \,,
\eeqn
where 
\beqn \label{pt12-app2}
     a_{1} = c_1, ~ \beta_{1}=b_1, ~d_{1}=c_2, ~ \mbox { and  for $k >1$:} \nonumber \\ 
     a_{k}=1, ~ 
     \beta_{k}=b_{k} z_{k-1} = b_{k} \sqrt{\frac{a_{k-1}}{a_{k-1} d_{k-1} - (\beta_{k-1})^2}}, 
     ~ d_{k}=c_{k+1}.
\eeqn  
In each stage, the quadratic system (\ref{quadratic-system}) has 4 solutions 
but for our purpose it is enough to consider only one among them
\beqn  \label{pt13-app2}
    x_k = \frac{1}{\sqrt{ a_{k} }}, ~ 
    y_k = -\frac{ \beta_{k} }{\sqrt{a_{k} \left[ a_{k} d_{k} - (\beta_{k})^2  \right]} }, ~
    z_k = \frac{\sqrt{a_{k}} }{ \sqrt{ a_{k} d_{k} - (\beta_{k})^2 }} \,.
\eeqn
Express the elements of the inverse matrix $S^{-1} = U  U^T$ respecting (\ref{elements-U}) 
\beqn \label{symbolic-inverse-elements}
    (S^{-1})_{i,j} &=& \sum_{k=1}^{K-1} U_{i,k} U_{j,k} 
        = \sum_{k=\max \{i,j\}}^{K-1} U_{i,k} U_{j,k}  \nonumber \\
   &=& \sum_{k=\max \{i,j\}}^{K-1}  (x_k)^2 (y_i \ldots y_{k-1})  (y_j \ldots y_{k-1}) z_{i-1} z_{j-1}    \,. 
\eeqn
Note that in (\ref{symbolic-inverse-elements}), recursion appears only in the 
expression of $\beta_k$: 
\beqn \label{pt14-app2}
   \beta_1 = b_1, ~ \beta_2 = b_2 \sqrt{\frac{c_1}{c_1 c_2 - b_1^2}} \,, \;
   \beta_{k} =  b_{k} \sqrt{\frac{1}{c_k - (\beta_{k-1})^2}},\; \mbox{ for } k >2\, . 
\eeqn


\end{document}